\documentclass[runningheads]{llncs}
\usepackage{graphicx}
\usepackage{booktabs}
\usepackage{multirow}
\usepackage[table,xcdraw,dvipsnames]{xcolor}
\usepackage{mathabx}
\usepackage{arydshln}

\usepackage{floatrow}
\newfloatcommand{capbtabbox}{table}[][]
\floatstyle{plaintop}
\restylefloat{table}

\newcommand{\bm}{BM25}

\newcommand{\nsbm}{$NS\textsubscript{\bm}$}
\newcommand{\nsrandom}{$NS\textsubscript{random}$}

\newcommand{\trec}{\texttt{TREC-DL}}
\newcommand{\qqp}{\texttt{QQP}}
\newcommand{\mantis}{\texttt{MANtIS}}

\newcommand{\ltr}{L2R}
\newcommand{\ls}{LS}
\newcommand{\tsls}{T-LS}
\newcommand{\wsls}{WSLS}
\newcommand{\tswsls}{T-WSLS}
\newcommand{\ns}{NS}

\newcommand{\bert}{\texttt{BERT}}
\newcommand{\withls}{w. \texttt{\ls}}
\newcommand{\withtsl}{w. \texttt{\tsls{}}}
\newcommand{\withtwsls}{w. \texttt{\tswsls{}}}

\newcommand{\minusp}{$\pm$}
\newcommand{\testup}{\blacktriangleup}
\newcommand{\testdown}{\blacktriangledown}
\newcommand{\testupsecond}{\smalltriangleup}
\newcommand{\testdownsecond}{\smalltriangledown}

\newcommand{\hiddenknowledge}{hidden similarity knowledge}

\begin{document}
\title{Weakly Supervised Label Smoothing}
%
%
\author{Gustavo Penha \and
Claudia Hauff}
\institute{TU Delft\\
\email{\{g.penha-1,c.hauff\}@tudelft.nl}}

\maketitle              
\begin{abstract}
We study Label Smoothing (\ls{}), a widely used regularization technique, in the context of neural learning to rank (\ltr{}) models. \ls{} combines the ground-truth labels with a uniform distribution, encouraging the model to be less confident in its predictions. We analyze the relationship between the non-relevant documents---specifically how they are sampled---and the effectiveness of \ls{}, discussing how \ls{} can be capturing ``\textit{\hiddenknowledge{}}" between the relevant and non-relevant document classes. We further analyze \ls{} by testing if a curriculum-learning approach, i.e., starting with \ls{} and after a number of iterations using only ground-truth labels, is beneficial. Inspired by our investigation of \ls{} in the context of neural \ltr{} models, we propose a novel technique called \emph{Weakly Supervised Label Smoothing} (\wsls{}) that takes advantage of the retrieval scores of the negative sampled documents as a weak supervision signal in the process of modifying the ground-truth labels. \wsls{} is simple to implement, requiring no modification to the neural ranker architecture. Our experiments across three retrieval tasks---passage retrieval, similar question retrieval and conversation response ranking---show that \wsls{} for pointwise BERT-based rankers leads to consistent effectiveness gains. The source code is available at \textcolor{RubineRed}{\url{https://anonymous.4open.science/r/dac85d48-6f71-4261-a7d8-040da6021c52/}}.
\end{abstract}

\section{Introduction}

Neural Learning to Rank (\ltr{}) models are traditionally trained using large amounts of strongly labeled data, i.e., human generated relevance judgements. For example, in ad hoc retrieval each instance is comprised of a query, a document and a relevance judgment. All the other documents in the collection that were not labeled as (non-)relevant for the query, while not specified explicitly, can be viewed as non-relevant for the query. Since utilizing an entire corpus for training a \ltr{} model is practically infeasible, the typical procedure is to rely on the top-$k$ ranked documents for a query obtained from an efficient (but less effective) retrieval model such as BM25. While research has shown that the negative sampler (\ns{}), i.e. the technique to select documents to use as negative samples for a query, matters a great deal in the effectiveness of the learned ranker~\cite{aslam2009document,li2019sampling,karpukhin2020dense,xiong2020approximate,cohen2019learning} there has been no work on how to make use of the \emph{scores} of the \ns{}, which are currently ignored in the training of \ltr{} models---only the content of the documents are employed. 

In this work we first aim to understand, in the realm of neural \ltr{}\footnote{Binary relevance prediction is quite different from other domains such as image classification and language modelling which employ up to thousands of distinct classes.}, a widely used and successful~\cite{zoph2018learning,zeyer2018improved,vaswani2017attention} regularization technique called Label Smoothing~\cite{szegedy2016rethinking} (\ls{}), that penalizes the divergence between the predictions and a uniform distribution. We begin by looking into how the choice of \ns{} impacts \ls{}, since in the binary relevance prediction problem \ls{} penalizes the model less than normal training when predicting a negative document as relevant and vice versa. We also analyze whether it is beneficial to use a curriculum-learning inspired procedure for the hyper-parameter that controls the \ls{} strength as shown by recent work on understanding \ls{} in other domains~\cite{dogan2019label,xu2020towards}. This initial exploration to understand \ls{} leads to the following research question: \textit{\textbf{RQ1} Is label smoothing an effective regularizer for neural \ltr{} (and if so, under what conditions)?} Our experimental results on three different retrieval tasks reveal that \ls{} is indeed an effective regularization technique for neural \ltr{}, specifically when \textbf{(a)} there is similarity between the relevant and the non-relevant sampled documents, i.e. when we use BM25 as the NS technique, and \textbf{(b)} a curriculum-like approach is used to control the strength of the smoothing.

Inspired by our findings, we propose the Weakly Supervised Label Smoothing (\wsls{}) technique which exploits the \ns{} retrieval scores, as opposed to \ls{} where all labels are smoothed equally, for training neural \ltr{} models. Instead of interpolating the ground-truth label distribution with a uniform distribution (as done in \ls{}), we interpolate it with the \ns{} score distribution. \wsls{} has two benefits compared to using the ground-truth labels: (a) it regularizes the neural ranker by penalizing overconfident predictions and (b) it provides additional supervision signal through weak supervision~\cite{dehghani2017neural} for the negative sampled documents. \wsls{} is simple to implement, and requires no modification to the neural ranker architecture, but only to the labels using weak supervision scores that are readily available. Our experiments to answer our second research question (\textit{\textbf{RQ2} Is \wsls{} more effective than \ls{} for training neural L2R models?}) reveal that \wsls{} is a better way of smoothing the labels by providing additional weak supervision obtained from the negative sampling procedure. We reach relative gains of 0.5\% in effectiveness across tasks.

\section{Background: Label Smoothing (\ls{})} \label{section:background}

Given an input instance $x$ (a query and document combination), two classes ($k=0$ means not relevant and $k=1$ relevant, and thus here $K=2$), a ground truth distribution $q(k \mid x)$ and predictions from the neural \ltr{} model $p(k \mid x)=\frac{\exp \left(z_{k}\right)}{\sum_{i=1}^{K} \exp \left(z_{i}\right)}$, where $z_{i}$ are the logits, we can use the cross entropy loss for training: $\ell=-\sum_{k=0}^{K} \log (p(k)) q(k)$, where \( q(k)=\delta_{k, y}, \) and \( \delta_{k, y} \) is Dirac delta (equals 1 for \( z=y \) and 0 otherwise). Maximizing the log-likelihood of the correct label is approached if the logit corresponding to the ground-truth label is much greater than all other logits: \( z_{y} \gg z_{k} \) for all \( k \neq y\). This encourages the model to be overconfident in its predictions, which might not generalize well. Label smoothing~\cite{szegedy2016rethinking} is a regularization mechanism to encourage the model to be less confident. Given a distribution \( u(k) \), \emph{independent} of the training example \( x, \) and a smoothing parameter \( \epsilon \), for a training example with ground-truth label \( y, \) we replace the label distribution \( q(k \mid x)=\delta_{k, y} \) with $q^{\prime}(k \mid x)=(1-\epsilon) \delta_{k, y}+\epsilon u(k)$. In \ls{} the uniform distribution is employed, i.e. $u(k)=1/K$.

While \ls{} is a widely used technique to regularize models, the reasons underlying its successes~\cite{zoph2018learning,zeyer2018improved,vaswani2017attention} and failures~\cite{kornblith2019better,seo2020self} remain unclear. M{\"u}ller et. al \cite{muller2019does} showed that while \ls{} impairs teacher models to do knowledge distillation\footnote{Knowledge distillation~\cite{hinton2015distilling} is the process of using the predictions of a teacher model with higher complexity/size as the ground-truth distribution for a weaker model.} it improves the models' calibration, i.e. how representative the predictions are with respect to the true likelihood of correctness~\cite{guo2017calibration}. 



\vspace{-0.5cm}
\subsubsection{\textbf{Curriculum Learning for Label Smoothing (\tsls{})}}
Xu et. al \cite{xu2020towards} argued that given the empirical evidence of \ls{} ineffectiveness in certain cases, it is natural to combine \ls{} with the ground-truth labels during training in a two-stage training procedure and thus proposed \tsls{}: starts training with \ls{}, i.e. $\epsilon>0$, and after $X$ training instances use normal training, i.e. $\epsilon=0$ (the unmodified ground-truth labels are used). Similarly, Dogan et. al \cite{dogan2019label} proposed to move from a distribution of labels smoothed by the similarity between label classes towards the ground-truth labels with a curriculum leaning procedure. In this paper we resort to \tsls{}\footnote{Initial experiments where we decreased $\epsilon$ linearly \cite{dogan2019label} were as effective as  \tsls{} \cite{xu2020towards}.}~\cite{xu2020towards} to test whether a curriculum learning inspired approach for $\epsilon$ is required or not in the training of neural \ltr{} models.

\vspace{-0.25cm}
\section{Weakly Supervised Label Smoothing (\wsls{})}
We propose to replace the uniform distribution $u(k)$ that is independent of the example $x$, with a weakly supervised function $w(k \mid x)$, which is readily available for documents with label $0$ as part of the negative sampling procedure of \ltr{}, at no additional cost: the negative sampler (NS) score\footnote{When building the training col. of triplets for DL TREC 2019/20, for each query there is 1 relevant passage; the non-relevant passages are retrieved using \bm{}.}. Specifically, $q^{\prime}(k \mid x)=(1-\epsilon) \delta_{k, y}+\epsilon NS(k \mid x)$, where $NS(k \mid x)$ is the negative sampling procedure score for instance $x$ and label class $k$. If we use $\bm{}$ to retrieve negative samples\footnote{Since the \bm{} scores are not between 0 and 1 we apply min-max scaling.}, then for $k=0$ we have $q^{\prime}(k \mid x)=(1-\epsilon) \delta_{k, y}+\epsilon \bm{}(x)$ and when $k=1$ we fall back to \ls{} since we have strong labeled data:  $q^{\prime}(k \mid x)=(1-\epsilon) \delta_{k, y}+\epsilon \frac{1}{K}$. In the same way we can induce a curriculum learning procedure for \ls{} resulting in \tsls{} (see \S \ref{section:background}), we can do it for \wsls{}, for which we refer to as \tswsls{}\footnote{This bears resemblance to strategies for combining strong and weak labeled data~\cite{shnarch2018will}.}.

\section{Experimental Setup}

\textbf{Tasks \& Datasets:} In order to evaluate our research questions, we resort to the three following retrieval tasks: passage retrieval using the 2020 Deep Learning track of TREC (\trec{}) dataset\footnote{Since the official test split is not available we split the dev. set into two.}, similar question retrieval with the Quora Question Pairs~\cite{iyer2017first} (\qqp{}) dataset and conversation response ranking with the \mantis{}~\cite{penha2019introducing} dataset. We use them due to the large amount of labeled examples (required for training neural ranking models) and diversity of tasks.

\noindent{\textbf{Implementation details \& Evaluation:} We use BERT-based ranking as a strong neural \ltr{} baseline. We follow previous research~\cite{lin2020pretrained} and fine-tune BERT using the \texttt{[CLS]} token to predict binary relevance---the query and the document are concatenated using the \texttt{[SEP]} token and used as input---using the cross-entropy loss and Adam optimizer~\cite{kingma2014adam} with $lr=5^{-6}$ and $\epsilon = 1^{-8}$. We train with a batch size of $32$ and fine-tune the models for 50000 training instances.We train and test each model 5 times using different random seeds with 10 total candidate documents by query\footnote{So for example in the passage retrieval task there are 10 passages and only one is relevant for each query and for similar question retrieval there are 10 questions.}. We resort to a standard evaluation metric in conversation response ranking~\cite{yuan2019multi,gu2020speaker}: recall at position $K$ with $n$ candidates\footnote{For example $R_{10}@1$ indicates the number of relevant documents found at the first position when the model has to rank 10 candidates.}: $R_n@K$.} Since all tasks here are concerned with re-ranking $R_n@K$ is a suitable sampled metric\footnote{A sampled metric uses a sample of documents instead of the whole collection~\cite{krichene2020sampled}.} to compare models on how high the relevant documents are ranked when having only $n$ candidates. We resort to a robust and widely used \ns{} to obtain such candidates: \bm{}. We refer to using the query as input to \bm{} and selecting the top $n-1$ ranked documents as \nsbm{}. We also use random sampling (\nsrandom{})---which samples candidate documents from the whole collection with the same probability and thus brings documents that are quite different from the relevant one---to better understand \ls{}.

\vspace{-0.5cm}
\begin{table}[]
\centering
\small
\caption{Average $R_{10}@1$ and the standard deviation results of 5 runs with different random seeds for \bert{} with label smoothing (\withls{}) and \bert{} with two-stage label smoothing (\withtsl{}) for different negative samplers during training (\nsbm{} and \nsrandom{}) and $\epsilon=0.2$ for the development set. Bold indicate the highest values for each dataset and $\testup{}$/$\testdown{}$ superscripts indicate significant gains and losses respectively over the baseline (\bert{}) using paired Student’s t-test with confidence level of 0.95.}
\label{table:label_smoothing}
\begin{tabular}{@{}lllllll@{}}
\toprule
 & \multicolumn{3}{c}{\nsbm{}} & \multicolumn{3}{c}{\nsrandom{}} \\ \cmidrule(l){5-7} \cmidrule(l){2-4}
 & \multicolumn{1}{c}{\trec{}} & \multicolumn{1}{c}{\qqp{}} & \multicolumn{1}{c}{\mantis{}} & \multicolumn{1}{c}{\trec{}} & \multicolumn{1}{c}{\qqp{}} & \multicolumn{1}{c}{\mantis{}} \\ \cmidrule(l){5-7} \cmidrule(l){2-4}
\bert{} & 0.568\minusp{}.00$^{}$ & 0.581\minusp{}.03$^{}$ & 0.612\minusp{}.01$^{}$ & \textbf{0.385}\minusp{}.01$^{}$ & 0.444\minusp{}.01$^{}$ & \textbf{0.350}\minusp{}.01$^{}$ \\ \hdashline
\withls{} & 0.564\minusp{}.01$^{\testdown{}}$ & 0.593\minusp{}.01$^{\testup{}}$ & 0.612\minusp{}.01$^{}$ & 0.304\minusp{}.05$^{\testdown{}}$ & 0.440\minusp{}.03$^{\testdown{}}$ & 0.348\minusp{}.01$^{\testdown{}}$ \\
\withtsl{} & \textbf{0.570}\minusp{}.01$^{\testup{}}$ & \textbf{0.598}\minusp{}.01$^{\testup{}}$ & 0.612\minusp{}.01$^{}$ & 0.382\minusp{}.02$^{\testdown{}}$ & 0.444\minusp{}.01$^{}$ & 0.345\minusp{}.01$^{\testdown{}}$\\ \bottomrule
\end{tabular}
\vspace{-1cm}
\end{table}

\section{Results}
\subsubsection{Effectiveness of Label Smoothing for Neural Ranking (RQ1)}
Table \ref{table:label_smoothing} displays the dev. set results\footnote{Since we do not do any hyper-parameter tuning for RQ1, we resort to the dev. set to avoid overusing the test set.} for the \ls{} and \tsls{} techniques when changing the \ns{}. The results reveal that when training \bert{} with \nsrandom{} to sample negative documents, it is not effective to use any type of label smoothing. In fact there is a consistent and statistically significant decrease in the effectiveness compared to \bert{}. In contrast, when we sample documents to train with \nsbm{} we observe that there are significant gains to train \bert{} with \tsls{}, with the exception of \mantis{} where there is no statistical difference. When we compare \ls{} with \tsls{}, we see that it is indeed beneficial to use a curriculum-learning approach for label smoothing (\tsls{}), which indicates that being more permissive of the mistakes in the first half of training is effective---this is in line with results obtained in other domains~\cite{dogan2019label,xu2020towards}. \textbf{This answers our first RQ positively: label smoothing is an effective regularization technique to train neural L2R models, with gains of 1\% of $R_{10}@1$ compared to standard training (\bert{}) on average across three different retrieval tasks when (a) using \nsbm{} and (b) a curriculum learning approach for \ls{}}. 

We hypothesize that label smoothing is effective for training neural \ltr{} models if the negative documents are similar to the relevant documents for the query. Our results when changing from \nsbm{} to \nsrandom{} support this hypothesis. Intuitively, if the negative document is random and thus very dissimilar to the query, using a label smoothing regularizer will penalize the model less for this mistake, which might hinder learning. When using label smoothing with a negative document that was sampled using \bm{}, we are penalizing the model less for choosing a document that is similar to the query in terms of exact matching words. In this way we are teaching the model the similarity between the classes relevant and non-relevant\footnote{A similar reasoning can be found in recent work which discusses that the similarity between classes on the wrong responses, i.e. ``\textit{\hiddenknowledge}"~\cite{hinton2015distilling}, is helpful for learning better neural networks~\cite{yuan2020revisiting,dogan2019label,furlanello2018born}.} by means of documents that are closer to the classification frontier. Our findings align with~\cite{mitra2017learning}: training with topically similar (but non-relevant) documents---as opposed to random documents---allows the model to better discriminate between documents provided by an earlier retrieval stage.

\subsubsection{Effectiveness of Weakly Supervised Label Smoothing (RQ2)}

\begin{figure}
\begin{floatrow}
\ffigbox[.40\textwidth]{%
    \includegraphics[width=.40\textwidth]{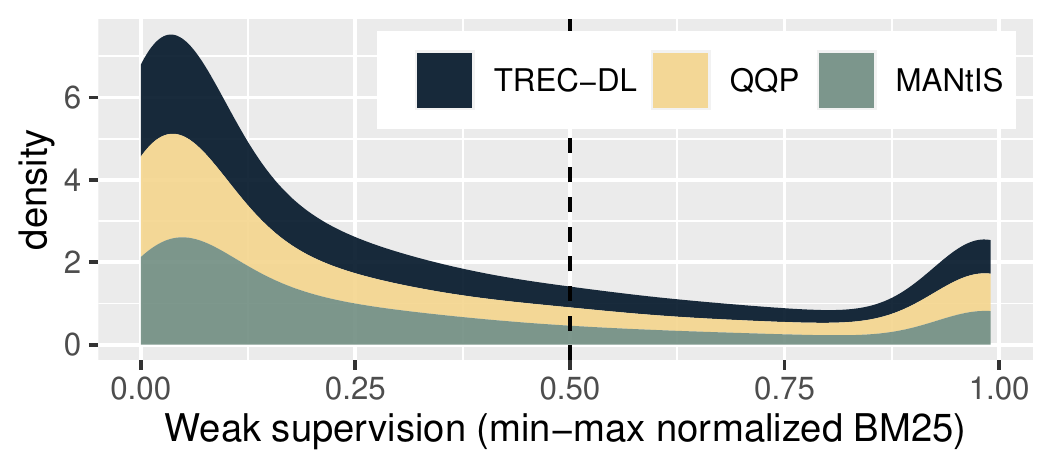}        
}{%
  \caption{Stacked and smoothed weak supervision distributions used for \wsls{} from the min-max normalized scores of \nsbm{}. The dashed vertical line indicates the distribution used by \ls{} (uniform with $K=2$).}%
  \label{fig:bm25_weak_supervision}
}
\capbtabbox[.55\textwidth]{%
\begin{tabular}{@{}llll@{}}
\toprule
 & \trec{} & \qqp{} & \mantis{} \\ \cmidrule(){2-4}
\bert{} & 0.599\minusp{}.00$^{}$ & 0.595\minusp{}.01$^{}$ & 0.609\minusp{}.01$^{}$ \\ \hdashline
\withtsl{} & 0.601\minusp{}.00$^{\testup{}}$ & 0.596\minusp{}.01$^{}$ & 0.607\minusp{}.01$^{}$ \\
\withtwsls{} & \textbf{0.604}\minusp{}.00$^{\testup{}\testupsecond{}}$ & \textbf{0.598}\minusp{}.01$^{\testup{}\testupsecond{}}$ & 0.609\minusp{}.01$^{\testupsecond{}}$ \\ \bottomrule
\end{tabular}
}{%
 \caption{Average $R_{10}@1$ and the standard deviation results of 5 runs with different random seeds for the test set. $\testup{}$/$\testdown{}$ and $\testupsecond{}$/$\testdownsecond{}$ superscripts indicate significant gains and losses over the baselines (\bert{}) and (\bert{} \withtsl{}) respectively using paired Student’s t-test with confidence level of 0.95 and Bonferroni correction.}%
 \label{table:test_results}
}
\end{floatrow}
\end{figure}


Before we dive into the effectiveness of \tswsls{}\footnote{Based on RQ1 results we use the two-stage approaches here (\tsls{} and \tswsls{}).} (RQ2), we investigate the distribution of the normalized weak supervision scores from \nsbm{} in Figure~\ref{fig:bm25_weak_supervision}. There is a high density for low scores indicating that only a few of the sampled documents receive scores close to the maximum of the list (0.99 score after min-max scaling) and most of them are closer to the minimum (0.00). This is very different from the uniform distribution used by \tsls{} (dashed vertical line), which does not change according to the sample, and with two classes ($K=2$) is equal to 0.5, whereas the mean of the weak supervision distribution is 0.33. This suggests that the optimal $\epsilon$ for \tswsls{} is different from \tsls{}.

\begin{figure}[]
\vspace{-0.5cm}
    \centering
    \includegraphics[width=0.8\textwidth]   {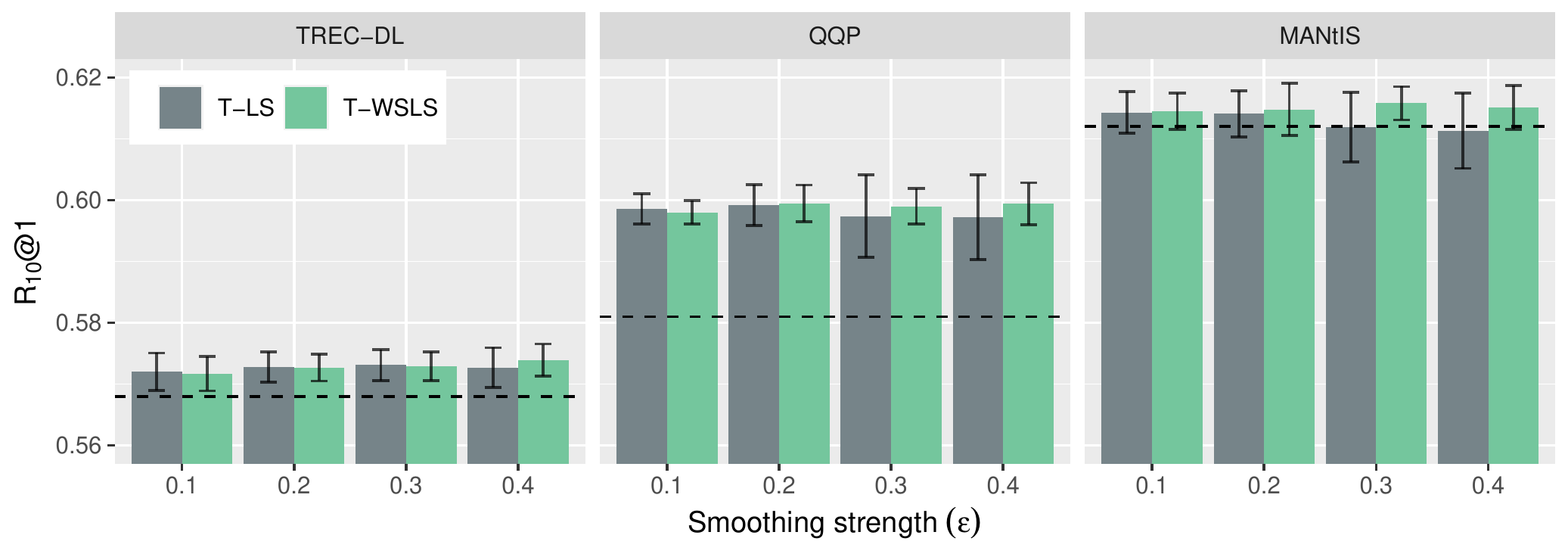}
    \setlength{\belowcaptionskip}{-15pt}
    \caption{\tsls{} and \tswsls{} sensitivity to the hyperparameter $\epsilon$ for the dev. set. Error bars indicate the 95\% confidence intervals for $R_{10}@1$ over 5 runs with different random seeds. Dashed horizontal lines indicate the baseline w/o label smoothing ($\epsilon=0$).}
    \label{fig:ls_vs_wsls}
\end{figure}

\definecolor{darkGreen}{HTML}{026e16}
Based on this observation, we test different values of $\epsilon$ on the dev. set in order to tune this hyper-parameter and use it on the test set. Figure \ref{fig:ls_vs_wsls} displays the effect of $\epsilon$ on the effectiveness of the proposed approach. The highest $R_{10}@1$ values are observed for \tswsls{}: 0.574 ({\color{darkGreen}$+1\%$} over the baseline w/o \tswsls{}) for \trec{} when $\epsilon=0.4$ , 0.600 ({\color{darkGreen}$+3.2\%$}) for \qqp{} when $\epsilon=0.2$ and 0.6151 ({\color{darkGreen}$+0.5\%$}) for \mantis{} when $\epsilon=0.4$. When we apply the best models (for both \tsls{} and \tswsls{}) found using the dev. set on the test set, we see in Table \ref{table:test_results} that \bert{} \withtwsls{} outperforms both \bert{} and \bert{} \withls{} with statistical significance (with the exception of \mantis{} where there is no difference). \textbf{This answers RQ2 indicating that \wsls{} is indeed more effective than \ls{} with statistically significant gains on all tasks against \tsls{} and with an average of 0.5\% improvement over \bert{}.}

\vspace{-0.9cm}
\section{Conclusion}
We studied LS in the context of neural \ltr{} models. Our findings indicate that \ls{} is effective when there is similarity between relevant and non-relevant documents and that using curriculum learning for the strength of the regularization is effective. We proposed a technique that combines the weak supervision scores of negative sampled documents with label smoothing (\wsls{}) which outperforms \ls{} on different retrieval tasks. In future work we will explore \wsls{} in a wider range of retrieval models and tasks. 

\bibliographystyle{splncs04}

\bibliography{references.bib}




\end{document}